\documentclass[reprint,superscriptaddress,showpacs, amsmath,amssymb, aps,
prl,floatfix,]{revtex4-1}
\usepackage{graphicx}
\usepackage{dcolumn}
\usepackage{color}
\usepackage{bm}
\newcommand{\eca}{EuCd$_2$As$_2$}
\newcommand{\ecs}{EuCd$_2$Sb$_2$}
\newcommand{\ecx}{EuCd$_2Pn_2$}
\usepackage[normalem]{ulem}
\graphicspath{{Images/}}
\begin{document}
	\preprint{APS/123-QED}
	\title{Resonant X-ray scattering study of diffuse magnetic scattering from the topological semimetals \eca\, and \ecs}
	\author{J.-R. Soh}
	\affiliation{Department of Physics, University of Oxford, Clarendon Laboratory, Parks Road, Oxford OX1 3PU, UK}%
	\author{E. Schierle}%
	\affiliation{Helmholtz-Zentrum Berlin fur Materialien und Energie, Albert-Einstein-Stra\ss e 15, D-12489 Berlin, Germany}%
	\author{D.\,Y.\,Yan}
	\affiliation{Beijing National Laboratory  for Condensed Matter Physics, Institute of Physics, Chinese Academy of Sciences, Beijing 100190, China}
	\author{H.\,Su}
	\affiliation{School of Physical Science and Technology, ShanghaiTech University, Shanghai 201210, China}
	\author{D. Prabhakaran}
	\affiliation{Department of Physics, University of Oxford, Clarendon Laboratory, Parks Road, Oxford OX1 3PU, UK}%
	\author{E. Weschke}%
	\affiliation{Helmholtz-Zentrum Berlin fur Materialien und Energie, Albert-Einstein-Stra\ss e 15, D-12489 Berlin, Germany}%
	\author{Y. F. Guo}
	\affiliation{School of Physical Science and Technology, ShanghaiTech University, Shanghai 201210, China}
	\author{Y. G. Shi}
	\affiliation{Beijing National Laboratory  for Condensed Matter Physics, Institute of Physics, Chinese Academy of Sciences, Beijing 100190, China}\author{A. T. Boothroyd}
	\email{andrew.boothroyd@physics.ox.ac.uk}
	\affiliation{Department of Physics, University of Oxford, Clarendon Laboratory, Parks Road, Oxford OX1 3PU, UK}%
	\date{\today}
	\begin{abstract}
		We have investigated the magnetic correlations in the candidate Weyl semimetals \ecx\, ($Pn$=As, Sb) by resonant elastic X-ray scattering (REXS) at the Eu$^{2+}$ $M_5$ edge. The temperature and field dependence of the diffuse scattering of \eca\, provide direct evidence that the Eu moments exhibit slow ferromagnetic correlations well above the N\'{e}el temperature.  By contrast, the diffuse scattering in the paramagnetic phase of isostructural \ecs\, is at least an order of magnitude weaker. The FM correlations present in the paramagnetic phase of EuCd$_2$As$_2$  could create short-lived Weyl nodes.
	\end{abstract}
	
	
\maketitle
One of the major themes in solid-state physics is the realisation of exotic types of relativistic electrons which travel at speeds much slower than the speed of light. These electrons, which mimic the behaviour of Weyl fermions, live in crystalline solids with very specific crystal structures and symmetry properties. In particular, there is a requirement for broken time-reversal symmetry or broken inversion symmetry, or a specific combination of the two~\cite{Hasan2017,Armitage2018,Burkov2018,Yan2017,Burkov2016}. The features in these Weyl semimetals (WSMs) which give rise to the exotic quasiparticles are the topologically-protected electronic band crossings called Weyl nodes. These come in pairs and have a definite chirality. 

Following the initial discovery of Weyl nodes in the TaAs structural family~\cite{Yang2015,Xu2015a,Lv2015,PhysRevLett.117.146401,Belopolski2015,Xu2015b,Xue1501092,PhysRevX.5.031023}, a wealth of other materials have been suggested to host Weyl fermions~\cite{Ruan2016b,Ruan2016,Yang2017,Wan2011,Chang2018,Liu2018,Borisenko2015,Liu2017b,Suzuki2016,Chang2016,PhysRevLett.117.056805,Wang2016eca}. However, some of these WSMs contain many pairs of Weyl nodes in the Brillouin zone, not always close to the Fermi level, making it is difficult to distinguish the contributions from individual nodes~\cite{Ruan2016b,Ruan2016,Yang2017,Wan2011,Chang2018,Liu2018,Borisenko2015,Liu2017b,Suzuki2016,Chang2016,PhysRevLett.117.056805,Wang2016eca}. Other WSMs have trivial (non-topological) bands crossing the Fermi energy which can obscure the topological effects arising from the Weyl nodes~\cite{Yang2017,Wan2011,Chang2018,Liu2018,Borisenko2015,Liu2017b,Suzuki2016,Chang2016,PhysRevLett.117.056805,Wang2016eca}.

There now a strong impetus to find new WSMs with a single pair of Weyl nodes (the minimum number permitted by symmetry~\cite{Armitage2018}) located close to the Fermi energy in an energy window not cluttered with other bands. Such a system would serve as a test bed for fundamental studies of Weyl physics. This desire for an ideal WSM has been expressed in the concluding chapters of several key review articles~\cite{Burkov2018,Yan2017,Hasan2017,Armitage2018}.

\begin{figure}[b!]
		\includegraphics[width=0.5\textwidth]{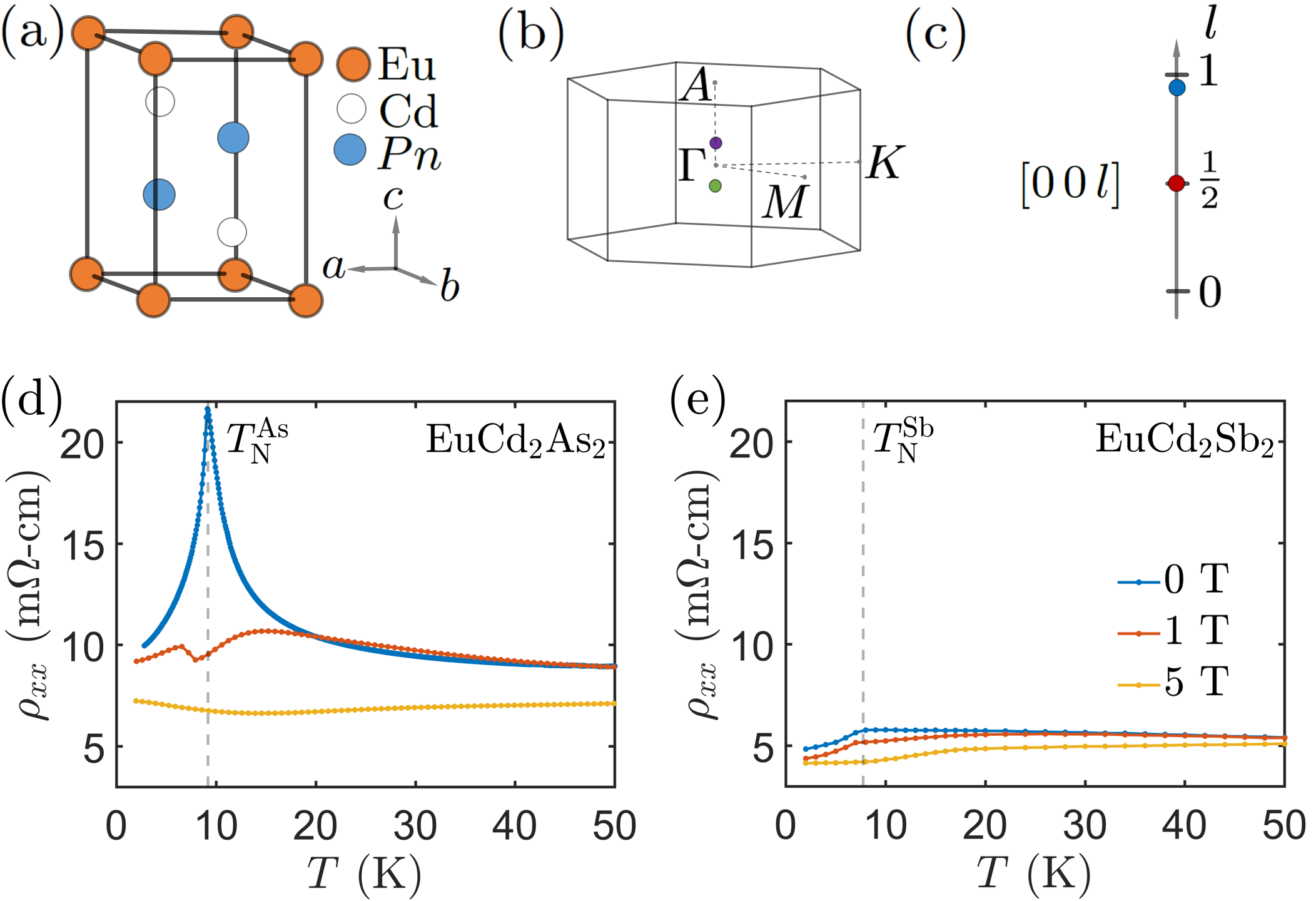}
		\caption{\label{fig:CB8a}(a) The unit cell of \ecx\, ($Pn$ = As, Sb) can be described by the $P\bar{3}m1$ space group. (b) In \eca\, the single pair of Weyl nodes, induced by magnetic exchange in a c-axis magnetic field, lie along the $A-\Gamma-A$ high symmetry line in the hexagonal Brillouin zone. (c) The REXS measurements were performed along $\Gamma - A$, which is the $(00l)$ direction in reciprocal space. (d) The in-plane resistivity peak of \eca\,at $T_\mathrm{N}^\mathrm{As}$ is fully suppressed in a field of 5 T. (e) On the other hand, isostructural \ecs\,  only displays a drop in  $\rho_{xx}$ below $T_\mathrm{N}^\mathrm{Sb}$. 
		}
	\end{figure}

Recently, it was predicted that an ideal WSM phase could be realised in \eca~\cite{Wang2019,Ma2019,Soh2019}. The single pair of Weyl nodes lie along the $A$--$\Gamma$--$A$ high-symmetry line in the hexagonal Brillouin zone [Fig.~\ref{fig:CB8a}(b)]. This finding has stimulated investigations into various physical properties of \eca, including anomalous conductance scaling~\cite{Behrends2019}, topological magnetotorsional effect~\cite{Liang2020}, Weyl-superconductor phase~\cite{Nakai2020}, quantum anomalous Hall effect~\cite{PhysRevB.99.235119}, axial anomaly generation~\cite{Hannukainen}, and non-reciprocal thermal radiation~\cite{Zhao}. The WSM state arises in \eca\, when the Eu moments are ferromagnetically (FM) aligned along the crystal $c$ axis. In zero field, \eca\, displays A-type anti-ferromagnetic (AFM) order below the N\'{e}el temperature ($T_\mathrm{N}^\mathrm{As} \simeq 9.5$\,K) with Eu moments lying in the $ab$ plane. A small coercive field of $H_c \simeq 2$\,T\, applied along the $c$ axis is sufficient to fully align the moments and create the WSM phase~\cite{Soh2019}. The Weyl nodes are created by exchange splitting of the conduction bands by magnetic coupling to the localised Eu $4f$ states.

Although an external magnetic field is required to induce the ideal WSM state in \eca, Ma \textit{et al.}~have proposed that spontaneous Weyl nodes can be detected in the paramagnetic (PM) phase of \eca\, without an applied field~\cite{Ma2019}. In this scenario, the Weyl nodes are induced by slow ferromagnetic (FM) correlations which are suggested to exist just above $T_\mathrm{N}^\mathrm{As}$. However, the $\mu_\mathrm{SR}$ technique employed in Ref.~\cite{Ma2019} to demonstrate the presence of cooperative spin fluctuations in the PM phase cannot distinguish between AFM and FM correlations, and only the latter can lift the two-fold degeneracy of the electronic bands necessary to create the Weyl nodes. 

	\begin{figure*}[t!]
		\includegraphics[width=0.9\textwidth]{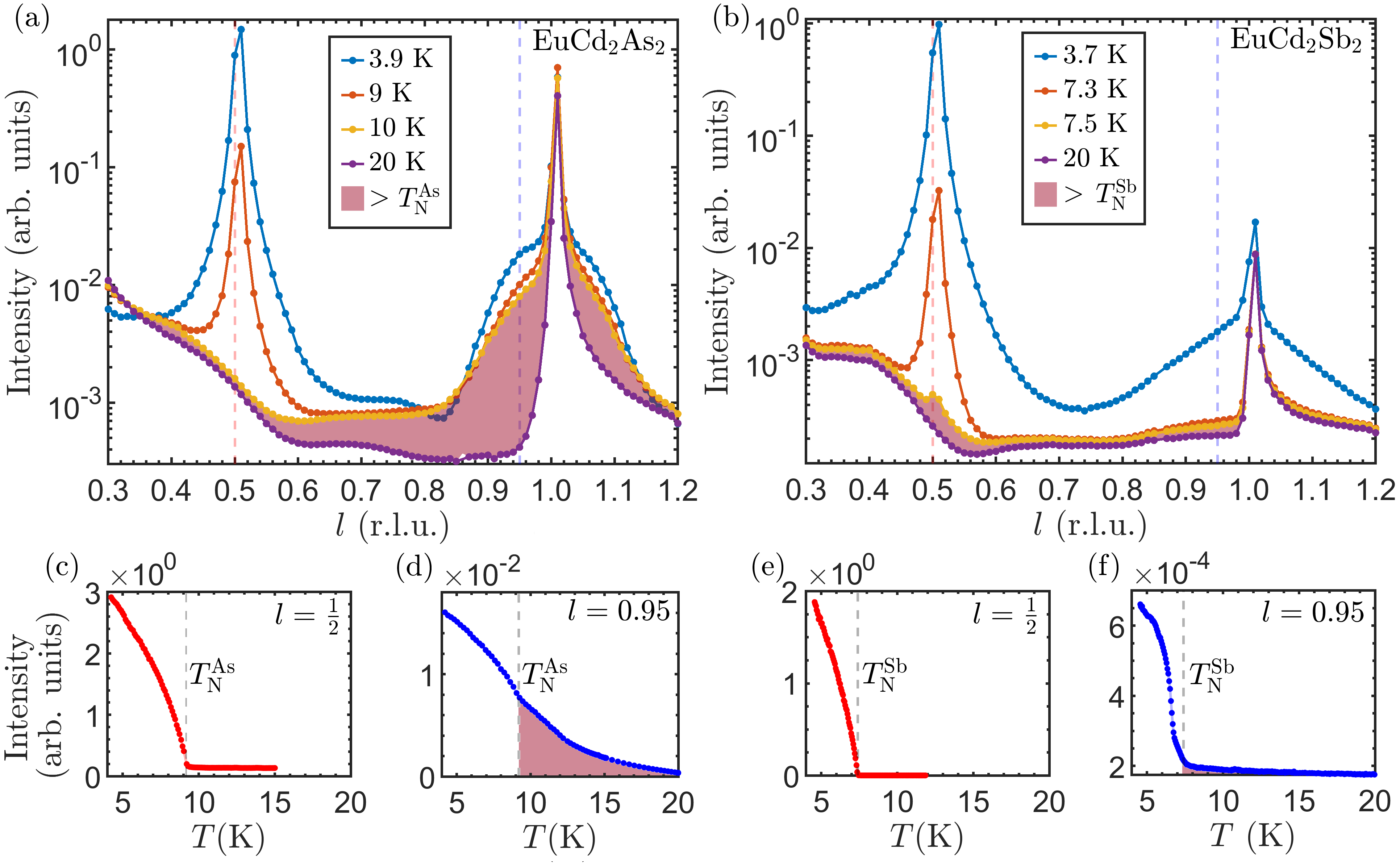}
		\caption{\label{fig:CB8a_Figures} (a) and (b) shows the temperature dependence of the REXS intensity with scattering vector \textbf{Q} along $(00l)$ in the range $0.3 \leq l \leq 1.2$ for \eca\, and \ecs, respectively. Cuts at $l=\frac{1}{2}$ and $l=0.95$ are shown in (c) and (d) for \eca, and (e) and (f) for \ecs. The shaded areas in (a), (b), (d) and (f) denote the changes in the scattered X-ray intensity between $T_\mathrm{N}^{Pn}$ and 20 K. (a) and (d) shows a development of a strong diffuse peak around the $(001)$ structural peak in \eca\, even before the onset of magnetic order at the $(00\frac{1}{2})$ AFM magnetic peak in (c). (b), (e) and (f) demonstrate that such a broad diffuse intensity is not present in \ecs.}
	\end{figure*}

Given that AFM and FM correlations in \eca\, have different characteristic wavevectors, the resonant elastic X-ray scattering (REXS) technique, which is wavevector-sensitive, offers a very direct way to separate the two types of correlation. The experiment can be done in practice by studying the diffraction signals at $\textbf{k}_\mathrm{AFM} = (00\frac{1}{2})$ and $\textbf{k}_\mathrm{FM} = (001)$, see Fig.~\ref{fig:CB8a}(c). Any diffuse scattering which is peaked at $\textbf{k}_\mathrm{FM}$  would signify FM correlations.  REXS is preferred over the more traditional technique of neutron diffraction because of the presence of the strongly neutron-absorbing elements Cd and Eu.

Here we use REXS to demonstrate decisively that FM correlations are present in the PM phase of \eca, in spite of the occurrence of AFM long-range order below $T_\mathrm{N}^\mathrm{As}$. We find that the diffuse magnetic scattering in the PM phase of \eca\, is at least an order of magnitude larger than that in isostructural \ecs.

Single-crystalline \eca\, and \ecs\, were grown by self-flux and chemical vapor transport methods, respectively, as described in Refs.~\cite{Schellenberg2011} and~\cite{Soh2018}. Comprehensive magnetization, magneto-transport and laboratory X-ray diffraction characterization of our samples has been reported in Refs.~\cite{Soh2019,Soh2018,Rahn2018} and are fully consistent with other reports on \ecx~\cite{Zhang2010,Su2019,Schellenberg2011,Wang2016,Jo2020,Ma2019,Junzhang2020}. 

For later reference we show in Figs.~\ref{fig:CB8a}(d) and (e) the temperature dependence of the longitudinal in-plane resistivity ($\rho_{xx}$) of \eca\, and \ecs, measured on a 14\,T  Physical  Property  Measurement System (Quantum Design) with the external magnetic field applied along the crystal $c$ axis. The $\rho_{xx}$ of \eca\, displays a sharp peak at $T_\mathrm{N}^\mathrm{As}$ which is fully suppressed in an applied field of 5\,T [Fig.~\ref{fig:CB8a}(d)]. In contrast, such a dramatic resistivity peak is absent in \ecs, and only a small drop in $\rho_{xx}$ at $T_\mathrm{N}^\mathrm{Sb}$ is observed [Fig.~\ref{fig:CB8a}(e)].

REXS was performed on the UE46-PGM1 beamline (BESSY II, Berlin) in the horizontal scattering geometry~\cite{ENGLISCH2001541}. To enhance the magnetic X-ray scattering from the Eu ions, the incident soft X-ray photon energy was tuned to the Eu M$_5$ edge at $\hbar\omega \simeq 1.1284$ keV (wavelength $\lambda=10.987$\AA) by a plane grating monochromator. Both crystals were mounted with the $a$- and $c$-axes in the horizontal scattering plane, and measurements were concentrated along the $(00l)$ line in reciprocal space. By good fortune, the $c$-axis lattice parameters of $c_\mathrm{As}$ = 7.29 \AA\, and $c_\mathrm{Sb}$ = 7.71 \AA\,  make  the scattering angles for the $001$ reflections close to 90$^\circ$ ($2\theta_\mathrm{As} = 97.8^\circ$ and $2\theta_\mathrm{Sb} = 90.8^\circ$), so we can suppress the Thomson scattering intensity of the strong  $001$ structural Bragg reflection by using photons with $\pi$ incident polarization.

We performed $(00l)$ scans of the REXS intensity in the range $0.3\leq l\leq 1.2$, which includes the AFM peak at ($00\frac{1}{2}$) and any diffuse magnetic scattering in the vicinity of the structural charge peak at ($001$). These zero-field measurements were performed at various temperatures up to $T = 20$\,K in the XUV diffractometer. Fixed-$l$ temperature dependent measurements of the scattered X-ray intensity were also made at $l = \frac{1}{2}$ and $l = 0.95$.
		
Subsequently, the \eca\, sample was transferred to the high-field diffractometer, where fixed-field temperature dependent $\omega$ scans at  $(00l)$ positions with $l = 0.943$ and $l = 1$ were performed at various applied field strengths up to $\mu_0H=0.5$\,T. Here, the $\omega$ scans are performed by rotating the crystal about an axis perpendicular to the $(h0l)$ scattering plane.

Figures~\ref{fig:CB8a_Figures}(a) and (b) plot (on a log scale) the zero-field measurements of the scattered X-ray intensity for \textbf{Q} along $(00l)$, at various temperatures. Both compounds exhibit a strong sharp peak centred on $(00\frac{1}{2})$ at $T < T_{\rm N}$ due to AFM order, and a sharp structural peak at $(001)$. In addition, diffuse scattering is observed under the $(001)$ peak indicating the presence of FM correlations. The diffuse scattering is present for both compounds, but with three important differences.  First, at the lowest temperature, $T\simeq 4$\,K, the diffuse scattering from \eca\, is at least one order of magnitude larger than that from \ecs. Second, the shape of the diffuse scattering is different, with prominent shoulders either side of the $(001)$ peak in \eca. Third, and most importantly, in \eca\, the diffuse scattering persists to temperatures above $T_{\rm N}$, whereas in \ecs\, it does not.

To further exemplify the salient features of the anti-ferromagnetic and the diffuse magnetic scattering, we plot the temperature dependence of the scattered X-ray intensity at $l = \frac{1}{2}$ and $l = 0.95$ in Figs.~\ref{fig:CB8a_Figures}(c) and (d) for \eca, and Figs.~\ref{fig:CB8a_Figures}(e) and (f) for \ecs. The onset of the AFM reflection is observed at $\textbf{Q} = (00\frac{1}{2})$ below  $T_\mathrm{N}^\mathrm{As}$ = 9.4\,K for \eca\,in Fig.~\ref{fig:CB8a_Figures}(c) and $T_\mathrm{N}^\mathrm{Sb}$ = 7.4\,K for \ecs\,in Fig.~\ref{fig:CB8a_Figures}(e). These N\'{e}el temperatures are consistent with earlier REXS studies~\cite{Soh2018,Rahn2018}. The shaded area in Fig.~\ref{fig:CB8a_Figures}(d) emphasizes that the REXS diffuse intensity for \eca\, extends up to at $\sim$20\,K, well above $T_\mathrm{N}^\mathrm{As}$. In contrast, the diffuse scattering from \ecs\, above $T_\mathrm{N}^\mathrm{Sb}$ is at least an order of magnitude weaker [Fig.~\ref{fig:CB8a_Figures}(f)]. 

The build-up of diffuse scattering around $(001)$ observed in \eca\, below $\sim$20\,K coincides with the onset of the $\rho_{xx}$ peak at $\sim$20\,K, see Fig.~\ref{fig:CB8a}(d). This suggests that the increase in resistivity in the PM phase of \eca\, on cooling towards $T_\textrm{N}^\textrm{As}$ can be attributed to charge carrier scattering from slow FM fluctuations of the Eu moments. This interpretation is also consistent with the magneto-resistive behaviour, Fig.~\ref{fig:CB8a}(d), which shows that the $\rho_{xx}$ peak in \eca\, can be fully suppressed in an applied magnetic field of $\sim$1\,T. The effect of the applied field is to fully align the Eu moments, reducing the contribution to the resistivity from FM fluctuations. In comparison, the $\rho_{xx}$ resistivity peak and magneto-resistive effect are much smaller in \ecs\, [Fig.~\ref{fig:CB8a}(e)], and the diffuse scattering signal  around $(001)$ is much weaker.

\begin{figure*}[t!]
	\includegraphics[width=\textwidth]{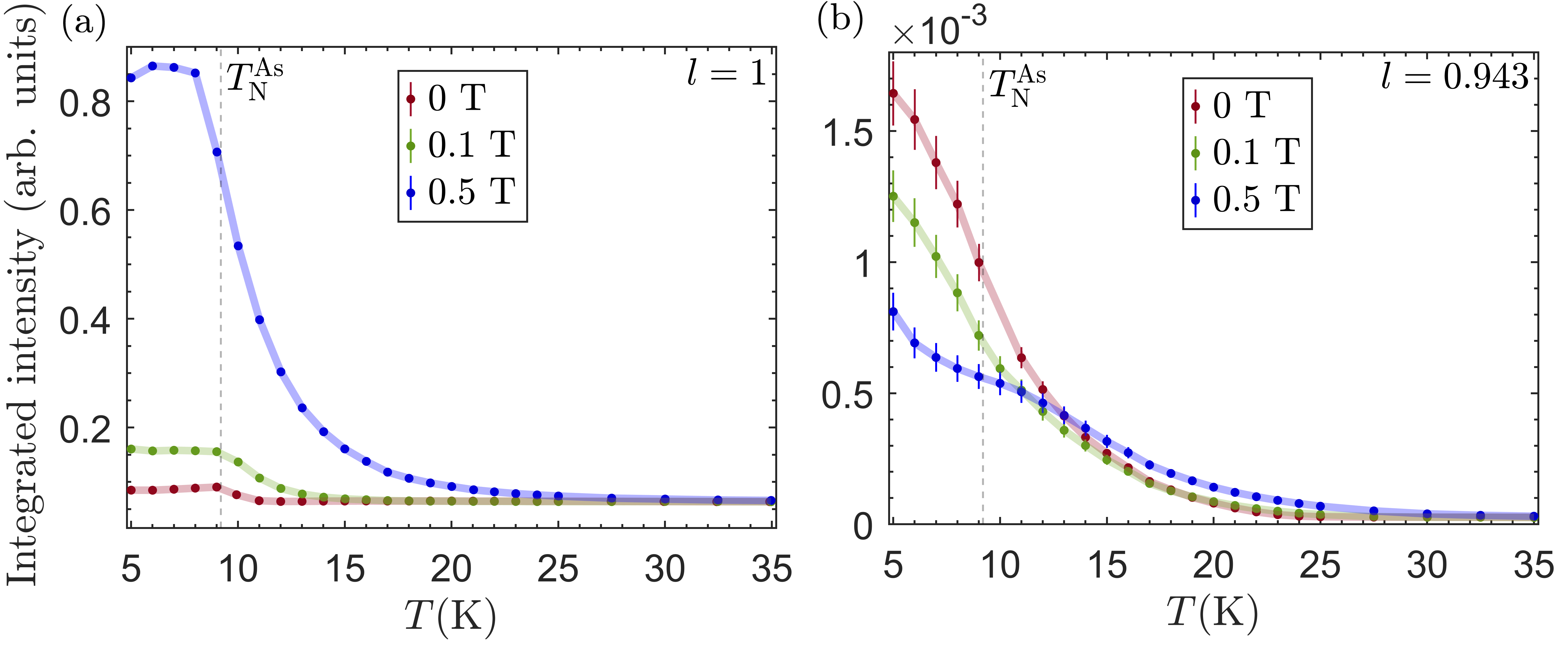}
	\caption{\label{fig:DR12c_HH0_Scan} Temperature dependence of the REXS intensity from \eca\, for various field strengths at $(00l)$ positions with (a) $l=1$, and (b) $l = 0.943$ . Peaks were measured in $\omega$-scans through $(00l)$, and the integrated intensity of the peaks is plotted.}
\end{figure*}

To augment the diffuse scattering data for \eca\, shown in Fig.~\ref{fig:CB8a_Figures}, we present in Fig.~\ref{fig:DR12c_HH0_Scan} measurements of the  temperature dependence of the REXS integrated intensity at two positions along $(00l)$ in applied magnetic fields up to 0.5\,T. The field was aligned along the scattered beam direction $\textbf{k}_f$, which was at an angle of approximately $45^\circ$ to the $c$ axis of \eca~\footnote{This field orientation was constrained by the limited scattering geometry permitted by the magnet.}.

Fig.~\ref{fig:DR12c_HH0_Scan}(a) plots the temperature dependence of the REXS intensity of the $001$ reflection. Upon cooling, we observe a monotonic increase of intensity with field and an almost temperature-independent plateau of intensity below $T_\textrm{N}^\textrm{As}$. This behaviour is consistent with Eu spins canting towards the applied field direction,  adding a magnetic component to the structural $(001)$ peak.

The diffuse scattered intensity at $(00l)$ with $l=0.943$ is shown in Fig.~\ref{fig:DR12c_HH0_Scan}(b). At this wavevector the scattering angle is $2\theta = 90^\circ$, so we expect purely magnetic contributions to the measured REXS intensity. For temperatures above $\sim$13\,K the intensity is enhanced by the field, but below $\sim$13\,K it is suppressed by the field. The $T>13$\,K behavior is consistent with an enhancement of FM correlations due to the field, while the crossover in behavior for $T<13$\,K  could be because as the temperature decreases the effectiveness of the  field to induce a FM moment increases, and so at low temperatures there is a greater transfer of intensity from the diffuse signal to the FM Bragg peak with field than at higher temperatures.

Based on the temperature and field dependence of the REXS measurements reported here, we find that cooperative magnetic fluctuations develop in \eca\, below $T \simeq 20$\,K and that they are related to the resistivity anomaly. Crucially, we have unambiguously shown that these magnetic fluctuations are associated with FM and not AFM correlations. If the FM correlations have a significant $c$-axis component then they could induce short-lived Weyl nodes which fluctuate along the $\Gamma$--$A$ line in the hexagonal Brillouin zone [see Fig.~\ref{fig:CB8a}(b)], as proposed by Ma \textit{et al.}~\cite{Ma2019}. 

In zero field, the FM diffuse scattering is still present below $T_\mathrm{N}$, both for \eca\, and \ecs. This may be due to $c$-axis stacking faults in the A-type AFM structure exhibited by both materials. As the Eu spins lie in the $ab$ plane in the AFM phase~\cite{Rahn2018}, it is not expected that the FM correlations present at $T < T_\textrm{N}$ would induce Weyl nodes. Diffuse scattering experiments that are sensitive to the direction of the spins would be needed to confirm this supposition.

It has recently been reported that slightly off-stoichiometric single crystals of \eca\, exhibit FM order below $T_{\rm C} \simeq 26$\,K, with the Eu spins lying in the $ab$ plane~\cite{Jo2020}. This finding, together with our observation of FM correlations in a sample which eventually develops AFM order, suggests that the magnetic propagation along the $c$ axis in \eca\, is delicately poised between FM and AFM order.

Given that the two isostructural compounds studied in this work have very similar magnetic ordering and lattice parameters, it is striking that they present such a significant difference in the diffuse REXS intensities above $T_\mathrm{N}$. We envisage that future work to determine the strength of the magnetic couplings between Eu moments could shed light on the origin of this disparity. The in-plane and inter-plane magnetic interactions could be obtained from the spin-wave spectrum measured by inelastic neutron scattering on samples containing isotopically-enriched europium and cadmium to reduce the the strong neutron absorption of the $^{151}$Eu and $^{113}$Cd isotopes. 

Finally, it is also instructive to consider other magnetic topological semimetals where magnetic fluctuations are predicted to significantly influence the topological features of the electronic bands near $E_\mathrm{F}$. Like \ecx, these related compounds also exhibit a rich interplay between magnetism and the topological charge carriers. For instance, the theoretical studies in Refs.~\cite{Zou2019,Tang2016} have  recently suggested that the magnetic fluctuations away from the fully ordered AFM configuration of Mn moments in the CuMnAs and CuMnP could open up an energy gap at the Dirac nodes, which are otherwise protected by the combination of the time-reversal and inversion symmetries. Similarly, the large longitudinal resistivity peak close to the Eu magnetic ordering temperature of the 112 pnictides EuMnSb$_2$ and EuMnBi$_2$~\cite{PhysRevB.100.174406,PhysRevB.96.205103,PhysRevB.98.161108,PhysRevB.90.075109,Borisenko2015,Masudae1501117}, may  point to the presence of FM fluctuations which could induce a WSM phase in the square Sb and Bi pnictide layers, respectively. Studies of the magnetic fluctuations in these and other compounds by diffuse scattering techniques like that presented in this work could advance our understanding of the role that magnetic fluctuations play in the topology of the bands in the wider family of magnetic topological semimetals.

To summarise, in this diffraction study we used X-rays tuned to the Eu $M_5$ absorption edge to reveal diffuse magnetic scattering in \ecx\, ($Pn = $As, Sb). The data  show conclusively that ferromagnetic correlations are present both above and below the antiferromagnetic ordering temperature in \eca, but only below the ordering temperature in \ecs.   The FM correlations in the paramagnetic phase of \eca\, could induce a transient WSM state above $T_{\rm N}$, consistent with the results of ARPES measurements \cite{Ma2019}.

We thank HZB for the allocation of synchrotron radiation beamtime. This work was supported by the U.K. Engineering and Physical Sciences Research Council (Grant Nos. EP/N034872/1 and EP/M020517/1), the National Natural Science Foundation of China (Grant No. 11874264), the starting grant of ShanghaiTech University, the Chinese National Key Research and Development Program (Grant No. 2017YFA0302901), the Strategic Priority Research Program (B) of the Chinese Academy of Sciences (Grant No. XDB33000000) and the EU Framework Programme for Research and Innovation HORIZON2020 (CALIPSOplus; Grant Agreement 730872). J.-R. Soh acknowledges support from the Singapore National Science Scholarship, Agency for Science Technology and Research.

\nocite{*}
\bibliography{library}
\end{document}